\UseRawInputEncoding
\documentclass[sn-mathphys,Numbered]{sn-jnl}

\usepackage{graphicx}%
\usepackage{multirow}%
\usepackage{amsmath,amssymb,amsfonts}%
\usepackage{amsthm}%
\usepackage{mathrsfs}%
\usepackage[title]{appendix}%
\usepackage{xcolor}%
\usepackage{textcomp}%
\usepackage{manyfoot}%
\usepackage{booktabs}%
\usepackage{algorithm}%
\usepackage{algorithmicx}%
\usepackage{algpseudocode}%
\usepackage{listings}%
\usepackage{bm}
\usepackage{times}
\usepackage{graphicx}
\usepackage{xspace}
\usepackage{siunitx}
\usepackage{xr}
\usepackage{lineno}
\usepackage{xcolor}
\newcommand{\apj}{Astrophys. J.}

\newcommand{\aap}{Astron. Astrophys.}
\newcommand{\apjl}{Astrophys. J. }
\newcommand{\mnras}{Mon. Not. R. Astron. Soc.}

\newcommand{\apjs}{Astrophys. J.}
\newcommand{\nat}{{\it Nature}}

\newcommand{\pasp}{Publ. Astron. Soc. Pac.}

\newcommand{\annrev}{Annu. Rev. Nuc. Part. Sci.}

\newcommand{\darkU}{Physics of the Dark Universe}

\newcommand{\sn}{SN\xspace}
\newcommand{\sne}{SNe\xspace}
\newcommand{\snia}{SN~Ia\xspace}
\newcommand{\sneia}{SNe~Ia\xspace}

\newcommand{\be}{\begin{equation}}
\newcommand{\ee}{\end{equation}}
\newcommand{\om}{\Omega_{\textrm{M}}}
\newcommand{\te}{\theta_{\rm E}}

\def\lsim{\raise0.3ex\hbox{$<$}\kern-0.75em{\lower0.65ex\hbox{$\sim$}}}
\def\gsim{\raise0.3ex\hbox{$>$}\kern-0.75em{\lower0.65ex\hbox{$\sim$}}}
\def\arcsec{\hbox{$^{\hbox{\rlap{\hbox{\lower4pt\hbox{$\,\prime\prime$}}
          }\phantom{\hbox{$\,\prime\prime$}}}}$}}
\def\farcs{\hbox{$^{\hbox{\rlap{\hbox{\lower4pt\hbox{$\,\prime\prime$}}
          }\phantom{\hbox{$\,\prime\prime$}}}}$}}

\def\arcmin{\hbox{$^{\hbox{\rlap{\hbox{\lower4pt\hbox{$\;\prime$}}
          }\hbox{$\frown$}}}$}}

\begin{document}
\title[]{Uncovering a population of gravitational lens galaxies with magnified standard candle SN Zwicky} 

\author*[1]{\fnm{Ariel} \sur{Goobar}}\email{ariel@fysik.su.se}
\author[1]{\fnm{Joel}   \sur{Johansson}}
\author[1]{\fnm{Steve} \sur{Schulze}}
\author[1]{\fnm{Nikki}  \sur{Arendse}}
\author[1]{\fnm{Ana}   \sur{Sagu\'es Carracedo}}
\author[2]{\fnm{Suhail} \sur{Dhawan}}
\author[1]{\fnm{Edvard} \sur{M\"ortsell}}
\author[3]{\fnm{Christoffer} \sur{Fremling}}
\author[3]{\fnm{Lin} \sur{Yan}}
\author[4]{\fnm{Daniel} \sur{Perley}}
\author[5]{\fnm{Jesper} \sur{Sollerman}}
\author[1]{\fnm{R\'emy} \sur{Joseph}}
\author[4]{\fnm{K-Ryan} \sur{Hinds}}
\author[3]{\fnm{William} \sur{Meynardie}}
\author[6,7,8]{\fnm{Igor} \sur{Andreoni}}
\author[9]{\fnm{Eric} \sur{Bellm}}
\author[10]{\fnm{Josh} \sur{Bloom}}
\author[11]{\fnm{Thomas E.} \sur{Collett}}
\author[3]{\fnm{Andrew} \sur{Drake}}
\author[3]{\fnm{Matthew} \sur{Graham}}
\author[3]{\fnm{Mansi} \sur{Kasliwal}}
\author[3]{\fnm{Shri R.} \sur{Kulkarni}}
\author[12]{\fnm{Cameron} \sur{Lemon}}
\author[13,14]{\fnm{Adam A.} \sur{Miller}}
\author[3]{\fnm{James D.} \sur{Neill}}
\author[15]{\fnm{Jakob} \sur{Nordin}}
\author[16]{\fnm{Justin} \sur{Pierel}}
\author[17]{\fnm{Johan} \sur{Richard}}
\author[18]{\fnm{Reed} \sur{Riddle}}
   \author[19]{\fnm{Mickael} \sur{Rigault}}
\author[20]{\fnm{Ben} \sur{Rusholme}}
\author[3]{\fnm{Yashvi} \sur{Sharma}}
\author[3]{\fnm{Robert} \sur{Stein}}
\author[10]{\fnm{Gabrielle} \sur{Stewart}}
\author[15]{\fnm{Alice} \sur{Townsend}}
\author[21,22]{\fnm{Yozsef} \sur{Vinko}}
\author[21]{\fnm{J. Craig} \sur{Wheeler}} 
\author[20]{\fnm{Avery} \sur{Wold}} 

\affil*[1]{\orgdiv{The Oskar Klein Centre, Department of Physics}, \orgname{Stockholm University}, \orgaddress{\street{Albanova University Center}, \city{Stockholm}, \postcode{SE-106 91}, \country{Sweden}}}

\affil[2]{\orgdiv{Institute of Astronomy and Kavli Institute for Cosmology}, \orgname{University of Cambridge}, \orgaddress{\street{Madingley Road}, \city{Cambridge}, \postcode{CB3 0HA},  \country{UK}}}

\affil[3]{\orgdiv{Cahill Center for Astrophysics}, \orgname{California Institute of Technology}, \orgaddress{\city{Pasadena}, \postcode{91125}, \state{CA}, \country{USA}}}

\affil[4]{\orgdiv{Astrophysics Research Institute}, \orgname{Liverpool John Moores University, IC2, Liverpool Science Park}, \orgaddress{\street{146 Brownlow Hill}, \city{Liverpool}, \postcode{L3 5RF},  \country{UK}}}

\affil[5]{\orgdiv{The Oskar Klein Centre, Department of Astronomy}, \orgname{Stockholm University}, \orgaddress{\street{Albanova University Center}, \city{Stockholm}, \postcode{SE-106 91},  \country{Sweden}}}

\affil[6]{\orgdiv{Joint Space-Science Institute}, \orgname{University of Maryland}, \orgaddress{\city{College Park}, \postcode{20742}, \state{MD}, \country{USA}}}
          
\affil[7]{\orgdiv{Department of Astronomy}, \orgname{University of Maryland}, \orgaddress{\city{College Park}, \postcode{20742}, \state{MD}, \country{USA}}}
         
 \affil[8]{\orgdiv{Astrophysics Science Division}, \orgname{NASA Goddard Space Flight Center}, \orgaddress{\street{Mail Code 661}, \city{Greenbelt}, \postcode{20771}, \state{MD}, \country{USA}}}

\affil[9]{\orgdiv{DIRAC Institute, Department of Astronomy}, \orgname{University of Washington}, \orgaddress{\street{3910 15th Avenue NE}, \city{Seattle}, \postcode{98195}, \state{WA}, \country{USA}}}

\affil[10]{\orgdiv{Department of Astronomy}, \orgname{University of California}, \orgaddress{\city{Berkeley}, \postcode{94720}, \state{CA}, \country{USA}}}

\affil[11]{\orgdiv{Institute of Cosmology and Gravitation}, \orgname{University of Portsmouth}, \orgaddress{\street{Dennis Sciama Building, Burnaby Road}, \city{Portsmouth},\postcode{PO1 3FX}, \country{UK}}}

\affil[12]{\orgdiv{Institute of Physics, Laboratoire d™Astrophysique}, \orgname{ Ecole Polytechnique Fédérale de Lausanne (EPFL)}, \orgaddress{\street{Observatoire de Sauverny}, \city{Versoix},\postcode{1290}, \state{CH}, \country{Switzerland}}}

\affil[13]{\orgdiv{Department of Physics and Astronomy}, \orgname{Northwestern University}, \orgaddress{\street{2145 Sheridan Road}, \city{Evanstone}, \postcode{60208}, \state{IL}, \country{USA}}}

\affil[14]{\orgdiv{Center for Interdisciplinary Exploration and Research in Astrophysics (CIERA)}, \orgname{Northwestern University}, \orgaddress{\street{1800 Sheridan Road}, \city{Evanstone}, \postcode{60201}, \state{IL}, \country{USA}}}

\affil[15]{\orgdiv{Institut fur Physik}, \orgname{Humboldt-Universität zu Berlin}, \orgaddress{\street{Newtonstra\ss Ÿe 15}, \city{Berlin}, \postcode{12489}, \country{Germany}}}

\affil[16]{\orgdiv{Space Telescope Science Institute},  \orgaddress{\city{Baltimore}, \postcode{21218}, \state{MD}, \country{USA}}}

 \affil[17]{\orgdiv{Univ Lyon, Univ Lyon1}, \orgname{Ens de Lyon, CNRS}, \orgaddress{\street{Centre de Recherche Astrophysique de Lyon UMR5574}, \city{Saint-Genis-Laval}, \postcode{F-69230}, \country{France}}} 
   
 \affil[18]{\orgdiv{Caltech Optical Observatorie}, \orgname{California Institute of Technology}, \orgaddress{\city{Pasadena}, \postcode{91125}, \state{CA}, \country{USA}}}

\affil[19]{\orgdiv{Universit\'e de Lyon}, \orgname{Universit\'e Claude Bernard Lyon 1, CNRS/IN2P3}, \orgaddress{\street{IP2I Lyon}, \city{Villeurbanne}, \postcode{F-69622}, \country{France}}} 

 \affil[20]{\orgdiv{IPAC}, \orgname{California Institute of Technology}, \orgaddress{\street{1200 E. California Blvd}, \city{Pasadena}, \postcode{91125}, \state{CA}, \country{USA}}}

\affil[21]{\orgdiv{Department of Astronomy}, \orgname{University of Texas at Austin}, \orgaddress{\street{2515 Speedway}, \city{Austin}, \postcode{78712-1205}, \state{TX}, \country{USA}}}

\affil[22]{\orgdiv{CSFK}, \orgname{Konkoly Observatory}, \orgaddress{\street{Konkoly Thege ut 15-17}, \city{Budapest}, \postcode{1121}, \country{Hungary}}}


\abstract{Detecting gravitationally lensed supernovae is among the biggest challenges in astronomy. It involves a combination of two very rare phenomena: catching the  transient signal of a stellar explosion in a distant galaxy and observing it through a nearly perfectly  aligned foreground galaxy that  deflects light towards the observer.
High-cadence optical observations with the Zwicky Transient Facility, with an unparalleled large field of view, led to the detection of a multiply-imaged Type Ia supernova (\snia), ``SN Zwicky", a.k.a. SN 2022qmx. Magnified nearly twenty-five times, the system was found thanks to the ``standard candle" nature of \sneia. High-spatial resolution imaging with the Keck telescope resolved four images of the supernova with very small angular separation, corresponding to an Einstein radius of only $\theta_E =\ang{;;0.167}$ and almost identical arrival times. The small $\theta_E$ and faintness of the lensing galaxy is very unusual, highlighting the importance of supernovae to fully characterise the properties of galaxy-scale gravitational lenses, including the impact of galaxy substructures.}

\maketitle

\section*{Main}

%
Our understanding of gravitational lensing due to the curvature of space-time, and the analogy with the deflection of light in optics, dates back to the work by Einstein in 1936 \cite{1936Sci....84..506E}. In this pioneering work he considered the case where both the lens and the magnified background source were stars in our Galaxy. Einstein concluded that the deflection angles were too small to be resolved with astronomical instruments. 
It was Zwicky \cite{1937PhRv...51..679Z} who one year later pointed out that if the source was extragalactic, entire galaxies or clusters of galaxies could be considered as gravitational deflectors. Hence, the image separation between multiple images of the source could be large enough to be resolved by astronomical facilities, as the size of the image separation scales with the lens mass and distance as 
the Einstein radius, $\theta_E \approx$\ang{;;0.9}$\left({M_{l} \over 10^{11} M_\odot}\right)^{1 \over2}\left( D_s \over {\rm 1~Gpc} \right)^{-{1 \over2}} \left( D_{ls} \over D_l\right)^{1 \over 2}$, where $M_\odot$ is the mass of the Sun, $M_l$  and $D_l$ are the lensing mass  and lens angular size distance, and $D_s$ and $D_{ls}$ are the distances from the observer to the source and between the lens and the source, respectively. 
\begin{figure}[!htb]
	\centering
	\includegraphics[width=\textwidth]{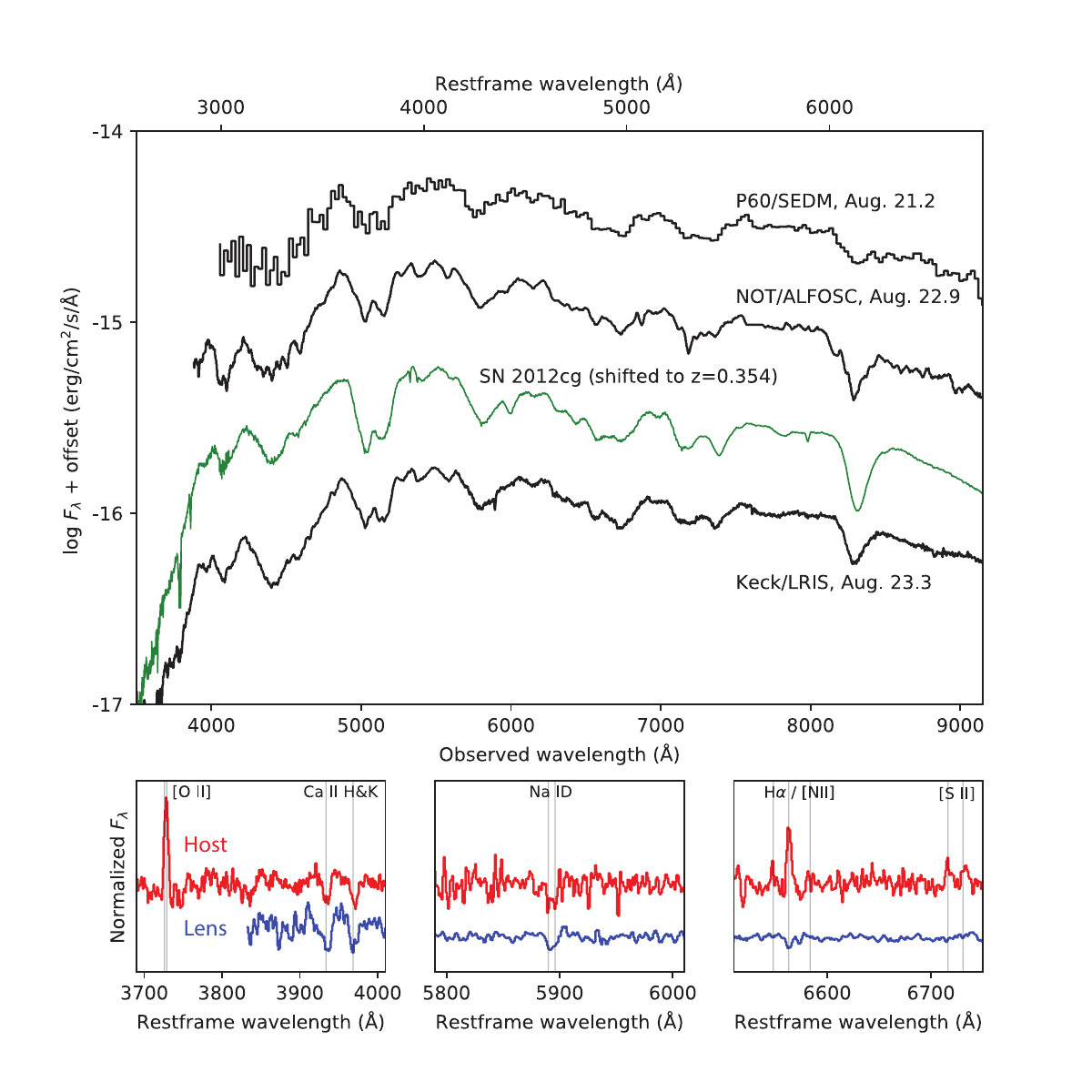}
	\caption{
	{\bf Spectroscopic identification of SN Zwicky as  a Type Ia supernova and redshift measurements of the \sn host galaxy and the intervening lensing galaxy.
	} 
The \sn spectra obtained with the P60, NOT and Keck telescopes (black lines) are best fit by a normal \snia spectral template.  The green line shows a comparison with the nearby Type Ia SN\,2012cg \cite{2015MNRAS.453.3300A} at a similar rest-frame phase, redshifted to $z=0.354$. The \sn flux peaked on August 17, 2022. Bottom panels show a zoomed-in view of a VLT/MUSE spectrum from 2022 September 30, displaying narrow absorption and emission lines, from which precise redshifts of the lens ($z=0.2262$) and host ($z=0.3544$) galaxies were determined. [O~{\sc ii}], Ca~{\sc ii} H \& K, Na~{\sc i d}, H${\alpha}$,  [N~{\sc ii}] and [S~{\sc ii}] lines can be seen at the rest-frame of the lens (blue lines) and host (red lines) galaxies.
\label{fig:spec}}
\end{figure}

 \subsection*{Strongly lensed supernovae in the era of wide-field time-domain surveys}
Besides the many observations of lensed galaxies and quasars, the feasibility to observe  strong gravitational lensing of explosive transients in the distant universe has only been demonstrated in recent years, see \cite{2019RPPh...82l6901O,2022ChPhL..39k9801L,2023arXiv230107729S} and references therein. \{PS1-10afx was the first highly magnified \snia  discovered. However, the lensing interpretation was made three years after the explosion \cite{2013ApJ...768L..20Q,Quimby2014}, by then the \sn it was too faint to resolve multiple images. Since then, the use of wide-field optical cameras in robotic telescopes at the Palomar observatory has led to notable breakthroughs. In \cite{2017Sci...356..291G} we reported the first discovery of a multiply-imaged \snia, iPTF16geu (SN 2016geu), by the intermediate Palomar Transient Factory (iPTF), a time-domain survey using a 7.3 sq.deg camera on the P48 (1.2-meter) telescope from 2013--2017. In 2018, a new camera was installed
 \cite{2020PASP..132c8001D} with a field of view of 47 sq.deg. The project, known as the Zwicky Transient Facility (ZTF) \cite{2019PASP..131a8002B,2019PASP..131g8001G}, has been monitoring the northern sky with a two- to three-day cadence in at least two optical filters for the past four years \cite{2019PASP..131f8003B}. The very large sky coverage makes ZTF well-suited to search for rare phenomena, such as gravitational lensing of \sne. On the other hand, the distance (redshift) probed by ZTF is limited by the relatively small mirror of the telescope, light pollution, non-optimal atmospheric conditions, and only having three optical filters at the P48 telescope. Furthermore, ZTF typically obtains image quality (angular resolution) of \ang{;;2} full-width half maximum (FWHM), and the camera has relatively large \ang{;;1} pixels. Hence, in most instances, it is  practically impossible to spatially resolve multiple-image systems with ZTF. Instead, the search for lensed sources makes use of the ``standard candle" nature of Type Ia supernovae, i.e., they have nearly identical peak luminosity. These explosions are used as accurate distance estimators in cosmology, which led to the discovery of the accelerated expansion of the universe, see \cite{2011ARNPS..61..251G} and references therein. 
 
 In addition to an imaging survey telescope, ZTF has access to a low spectral-resolution  integral field spectrograph, SEDM \cite{2018PASP..130c5003B}, on the neighboring 1.5-meter telescope at Palomar (P60), used to spectroscopically classify about ten \sne every night as part of the Bright Transient Survey (BTS), where transients brighter than $19$ mag are classified within timescales of a few days, aiming to obtain $>95\%$ spectroscopic completeness to 18.5 mag or brighter \cite{2020ApJ...895...32F}. 
 Besides providing the classification of the transients, the SEDM spectrum is used to measure the SN redshift. 

 \subsection*{The discovery of SN Zwicky}
 Lensed system candidates are selected by ZTF for further spectroscopic screening when a \snia is found at a redshift above $z=0.2$, where there is essentially negligible sensitivity for detection in BTS, unless the SN is greatly magnified by an intervening deflector. This was the case for ``SN Zwicky" (a.k.a. ZTF22aaylnhq and SN 2022qmx), located at right ascension $17^h$$35^m$$44.32^s$  and declination \ang{+04;49;56.90} (J2000), 
where an SEDM spectrum from 2022 August 21 showed it to be a Type Ia SN at $z=0.35$, as shown in Fig.~\ref{fig:spec}. At that point we alerted the \sn community of the discovery of a lensed \snia \cite{2022TNSAN.180....1G}. 

Spectroscopic observations following the time-evolution of the SN were carried out using multiple facilities, the 2.56-meter Nordic Optical Telescope at the Canary Islands, the Keck observatory in Hawaii, the 11-meter Hobby-Eberly Telescope at the McDonald Observatory in Texas and at ESO's 8-meter Very Large Telescopes (VLT) at the Paranal Observatory in Chile. In particular, multiple narrow emission and absorption lines of the SN host galaxy were found with LRIS/Keck and MUSE/VLT, refining the source redshift to $z=0.3544$, as shown in the bottom panels of Fig.~\ref{fig:spec}. As the \sn faded and the foreground galaxy spectral energy distribution (SED) became more prominent, the Ca~II doublet $\lambda \lambda 3933, 3968$ was found in absorption lines redshifted to $z=0.22615$, the location of the deflecting galaxy.

\subsection*{Follow-up observations}
The discovery from ZTF was also followed-up with high-spatial resolution instruments.
Observations with laser guide star enhanced seeing at the Very Large Telescope with the HAWKI imaging camera in the near-IR, and optical spectrophotometry with MUSE, reduced the point spread function (PSF) width to about \ang{;;0.4}. However, this was  still not enough to resolve the system. On 2022 September 15, 
multiple images of the system were first resolved at near-IR wavelengths at the Keck telescope, using the Laser Guide Star aided Adaptive Optics (LGSAO) with the Near-IR Camera 2 (NIRC2) \cite{2022TNSAN.194....1F}, yielding an image quality of \ang{;;0.09}
FWHM in the $J$-band  centered at 1.2 $\mu$m,  shown in Fig.~\ref{fig:zoom}, where the four \sn images are labeled A-D. 

On September 21, following our announcement of the discovery \cite{2022TNSAN.180....1G}, a previously approved program aimed to target lensed \sne by the LensWatch collaboration resolved the multiple images of SN Zwicky using the optical filters $F475W$, $F625W$ and $F814W$ (where the names correspond to the approximate location of the central wavelength in nanometers) on the UVIS/WFC3 Camera on the Hubble Space Telescope (HST) \cite{2022TNSAN.196....1P}. {A detailed description of the HST observations of SN Zwicky is presented in \cite{2022arXiv221103772P}.}
\begin{figure}[!htb]
	\centering
	\includegraphics[width=\textwidth]{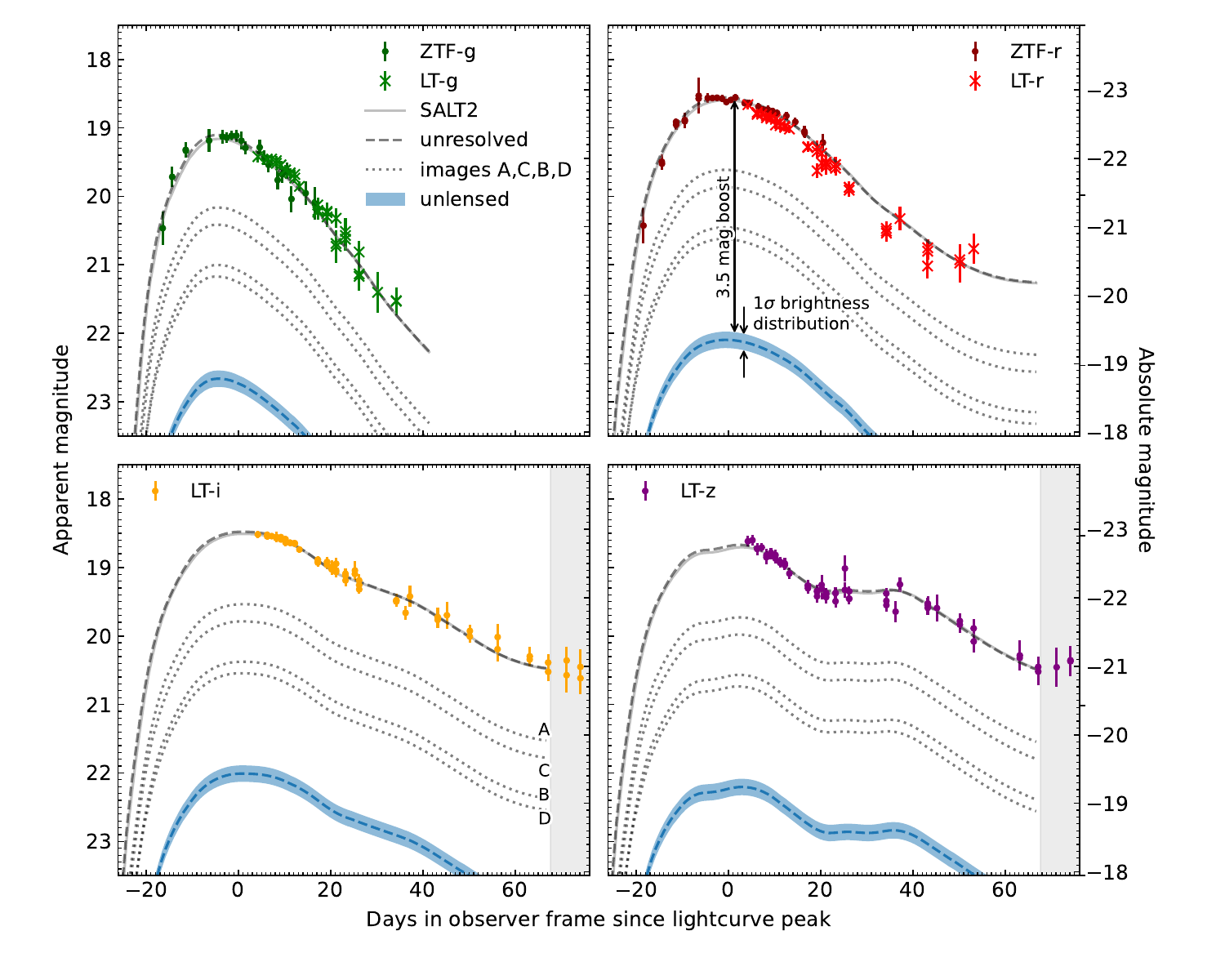}
	\caption{%
	{\bf Multi-color lightcurve of SN Zwicky showing that the supernova is 3.5 magnitudes 
	brighter than an unlensed SN at the same redshift.}
	The magnitudes are measured with respect to time of maximum light (Modified Julian Date 59808.67, 
 corresponding to August 17, 2022) in ZTF $g$ and $r$ and Liverpool Telescope $griz$ filters. 
		The solid lines show the best fit SALT2 \cite{Guy:2007js} model to the spatially unresolved data. The blue dashed lines indicate the expected lightcurves at $z=0.354$ (without lensing) where the bands represent the standard deviation of the brightness distribution for \sneia.  In order to fit the observed lightcurves, a brightness increase corresponding to 
		$3.5$~magnitudes is required. Also shown as dotted lines are the modeled individual lightcurves for the four \sn images A--D. The flux ratios were measured from the NIRC2 data in Fig.~\ref{fig:zoom}. From these lightcurves we extract the time-delays between the images, all in units of days,  $\Delta t_{AB} = -0.4 \pm 2.9$, $\Delta t_{AC} =-0.1 \pm 2.3$, and $\Delta t_{AD} = -0.1 \pm 2.7$, as described in the Supplementary Material section. The shaded areas in the lower panels indicate the regions with data outside the phase range where the SALT2 model is defined, therefore, excluded from the lightcurve fit analysis. {The error bars correspond to 1 standard deviation.} 
  \label{fig:lc}
  }
\end{figure}

\subsection*{Results}
Fig.~\ref{fig:lc} shows the unresolved photometric ground-based observations collected at P48 
and the Liverpool telescope on the Canary Islands between 2022 August 1 and October 30. These were  used to estimate the peak flux and lightcurve properties of the \sn with the SALT2 
lightcurve fitting tool \cite{Guy:2007js}, including corrections for lightcurve shape and colour excess given the \sn redshift, as  well as the extinction in the Milky-Way in the direction of the \sn.
Furthermore, the four resolved \sn images were used to explore the possibility of additional reddening by dust in the lensing galaxy. Unaccounted dimming of light would lead to an underestimation 
of the lensing amplification. The HST and NIRC2 observations  for each \sn image were compared with the \snia spectral template from \cite{hsiao2007} allowing for possible differential dust extinction in the lens following the reddening law in \cite{cardelli1989}. 

\begin{figure}[!htb]
\centering
	\includegraphics[width=0.6\textwidth]{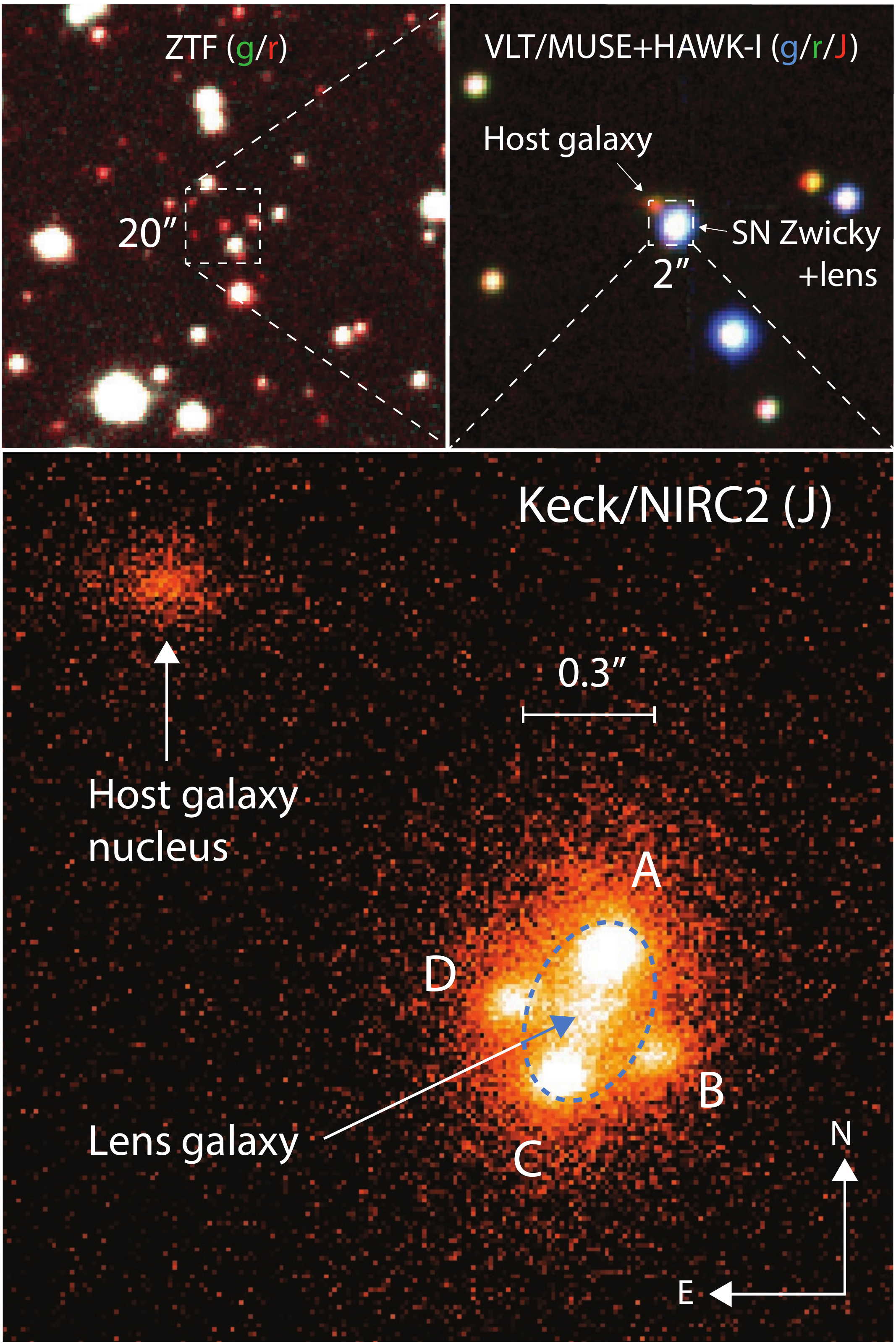}
	
	\caption{%
	{\bf Image of the field of SN Zwicky using pre-explosion $\boldsymbol{g}$ and $\boldsymbol{r}-$band images from ZTF. The top right panel shows a composite image using VLT MUSE ($\boldsymbol{g}$/$\boldsymbol{r}$) and HAWK-I $\boldsymbol{J}$-band data. The bottom panel shows the Keck/NIRC2 $\boldsymbol{J}$-band image where the four \sn images are resolved.}
	The top left panel shows a $\ang{;2;}\times \ang{;2;}$ section of the ZTF $g$ and $r$-band pre-SN images (FWHM \ang{;;2.3}), centered on the location of SN Zwicky. Top right panel shows a zoomed-in composite image of SN Zwicky using AO enhanced VLT MUSE and HAWK-I observations on September 2 and 4 (FWHM $\sim \ang{;;0.4}$).
	Bottom panel shows a $\ang{;;2} \times \ang{;;2}$ portion of the Keck/NIRC2 LGSAO $J$-band image (FWHM \ang{;;0.09}), resolving the four multiple images of SN Zwicky (labelled A, B, C, D). The blue dashed ellipse shows the critical line of the lens, corresponding to the inferred Einstein radius $\theta_E =\ang{;;0.167}$ (0.6 kpc at $z=0.2262$), enclosing the lens 
	$M=(7.82 \pm 0.06)\cdot 10^{9}\,M_\odot$. The host galaxy nucleus is located \ang{;;1.4} to the northeast of the lens, implying that SN Zwicky exploded at a projected distance of 7 kpc from the centre of its host galaxy.
		\label{fig:zoom}
		}
\end{figure}

No evidence for differential extinction between the different images was found. The lightcurve fit model included the four individual \sn images, described by the SALT2 model with arbitrary time delays, but otherwise sharing the same lightcurve shape parameters, $x_1$ and $c$. The time delays were constrained by a prior on the image flux ratios from the NIRC2 observations shown in Fig.~\ref{fig:zoom}.        
We find $\Delta t_{\rm AB} = -0.4 \pm 2.9$, $\Delta t_{\rm AC} = -0.1 \pm 2.3$, and $\Delta t_{\rm AD} = -0.1 \pm 2.7$ (in units of days), where the indices A-D refer to the \sn images in Fig.~\ref{fig:zoom}. The resulting lightcurve fit is shown in Fig.~\ref{fig:lc} and compared to the SALT2 model. The small time-delays are also consistent with the spectra of SN Zwicky shown in Fig.~\ref{fig:spec} being at a single \sn phase.

The fitted  SN model parameters, i.e. lightcurve shape and colour from the SALT2 model are 
{$x_1 = 1.083 \pm 0.094$ and $c = -0.007 \pm 0.007$}. {The lack of colour excess confirms that differential extinction is negligible. Since the lightcurve parameters errors do not include the model covariance, we conservatively add the SALT2 model error floor of $\sigma(x_1)=0.1$ and $\sigma(c)=0.027$ mag \cite{guy2010} in quadrature to the fit errors. 
Using the inferred apparent magnitude and the SALT2 parameters above},   we find a total magnification of $\Delta m = -3.44 \pm 0.14$ mag, assuming standard cosmological parameters from \cite{Planck2020:cosmo} and restframe $B$-band  \sneia  peak absolute magnitude of $-19.4$ mag for the average \snia lightcurve width and colour, and intrinsic brightness scatter of 0.1 mag. Since the inferred stellar mass of the host galaxy is $M_\star \lsim 10^{10} M_\odot$, mass-step corrections for the \snia absolute magnitude  as suggested in \cite{2010MNRAS.406..782S} are not required. 
 In summary, we find that including the four images, SN~Zwicky is {$23.7 \pm 3.2$} times brighter than {the observed flux of} normal \sneia at {the same } redshift, after applying colour and lightcurve shape corrections.


The Keck NIRC2 $J$-band image was used to obtain a lens model to account for the { observed \sn image positions, irrespectively of their fluxes. 
Assuming a singular isothermal ellipsoid \cite{1993LIACo..31..571K,1994A&A...284..285K} for the lens potential, we report an ellipticity $\epsilon_e=0.35\pm 0.01$ and Einstein radius $\theta_{\textrm{E}}=0.1670~\pm$~\ang{;;0.0006}. The mass enclosed within the ellipse with semi-major axis 0.78 kpc and semi-minor axis 0.51 kpc is $M=(7.82 \pm 0.06)\cdot 10^{9}\,M_\odot$. The lens model predictions for the time delays are in excellent agreement with the fitted values from the lighturves in Fig.\ref{fig:lc}.
Further details regarding the lens modelling are presented in the {Methods Section}.}

{Interestingly, the individual image magnifications predicted for SN Zwicky by the smooth macro lens model are inconsistent with the observed flux ratios. According to the lensing model, the observed fluxes of the \sn images A and C are factors of $>4$ and $>2$ too large, respectively, compared to images B and D.
 Given the small time-delays, this discrepancy cannot be accounted by different phases between the \sn images. Other explanations need to be considered, e.g., excess magnification and demagnification from milli- and microlensing effects arising from stars and substructure in the lens galaxy \cite{1989Natur.338..745K,1994ApJ...429...66W}. Since microlensing effects are capable of perturbing magnifications without altering image locations, these were also put forward to explain the observed flux ratios of iPTF16geu \cite{diego2022}, displaying differences between the observed and modelled image flux ratios of similar magnitude.}
Probing microlensing in these central regions opens a new window to directly measure the central stellar initial mass function (IMF) \cite{2018MNRAS.478.5081F} and test claims that the IMF may be heavier in the centres of galaxies \cite{2017ApJ...841...68V}.  
As detailed in Methods Section, the lack of time dependence in the image flux ratios, or anomalous variation in the unresolved lightcurves, gives a lower limit for the substructure masses of 0.02 $M_\odot$, if the discrepancy from the smooth lens model is caused by microlensing. From the lack of further image splitting of the four individual \sn images,  we infer an upper limit for the substructure mass of $3\cdot 10^7 M_\odot$.  

In order to check the impact of added macro lens model complexity, we have studied cases where the lens mass distribution is modelled with two matter components; one where the surface mass density follows the lens light distribution (with an arbitrary mass to light ratio, possibly interpreted as a baryonic mass component), and a second one introducing a dark matter halo with additional flexibility on density profiles. In spite of the added extra complexity, the quality of the fit to the \sn image positions does not improve, and only induces shifts in the predicted flux ratios below 5 \%, i.e., very small in comparison with the observed flux ratio anomalies.

 \begin{figure}[!htb]
\centering
	\includegraphics[width=0.99\textwidth]{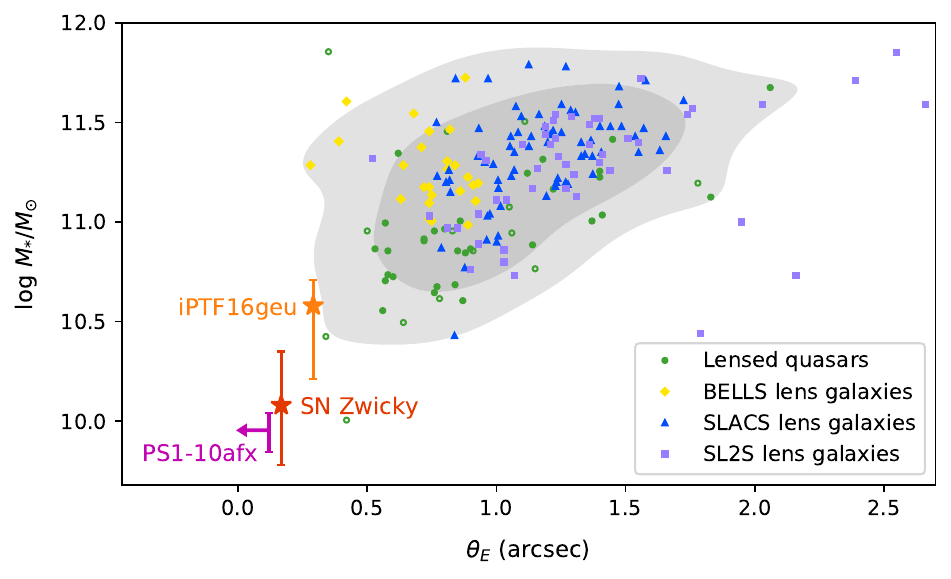}
	
	\caption{%
	{\bf Stellar mass versus Einstein radius for lens galaxies discovered in galaxy surveys, demonstrating that SN Zwicky, iPTF16geu, and the unresolved lensed supernova PS1-10afx} point to a poorly known population of small image separation lensing systems.
    Strongly lensed galaxy systems are represented by yellow diamonds for the BELLS sample \cite{Brownstein2012}, blue triangles for SLACS lenses \cite{Auger2009} and purple squares for SL2S lenses \cite{Sonnenfeld2013}. The green dots correspond to lensed quasars \cite{Oguri2014}, of which the filled dots have been detected optically and the open dots through radio emission. The shaded grey contours show the $90\%$ and $68\%$ confidence levels for the full sample of {155} lensed galaxies and {45} lensed quasars. {The lensed supernova data are presented as median values $\pm$ one standard deviation.} For the unresolved lensed supernova PS1-10afx, only an upper limit of the Einstein radius is available \cite{Quimby2014}. 
    The stellar mass derivation for SN Zwicky is detailed in the Methods Section.
	\label{fig:thetavsmass}
		}
\end{figure}

The demonstrated ability to discover multiply-imaged \sne makes it feasible to accomplish Refsdal's pioneering proposal \cite{1964MNRAS.128..307R} to use time-delays for strongly lensed \sne to measure the Hubble constant. This will require systems with time-delays of several days, i.e., longer than for SN Zwicky. 
The small physical scale of the lens probed by SN Zwicky, as well as iPTF16geu \cite{2017Sci...356..291G}, make these \sne unique tracers for uncovering a population of systems that otherwise would remain undetected, as shown in Fig.~\ref{fig:thetavsmass}, probing the mass distribution of the central, densest, regions of lensing galaxies. 
Both of these multiply-imaged \sneia were identified without preselections on e.g., association with bright galaxies or clusters, emphasizing the importance of untargeted surveys for unexpected discoveries.   





\newcommand{\Zw}{SN Zwicky\xspace}

\section*{Methods}

\subsection*{Supernova survey and follow-up} 
The Zwicky Transient Facility at Palomar Observatory is monitoring the transient sky at optical wavelengths since 2018 \cite{2020PASP..132c8001D,2019PASP..131a8002B,2019PASP..131g8001G,2019PASP..131a8003M}. `SN Zwicky" (a.k.a. SN\, 2022qmx) \cite{2022TNSAN.180....1G} was discovered under the Bright Transient Survey (BTS) program \cite{2020ApJ...895...32F}. The first detection of the SN was in a ZTF $g$-band image from 2022 August 1. It was saved to the BTS as a SN candidate by on-duty scanners on August 3 \cite{2022TNSTR2198....1F} and subsequently assigned to the queue for spectroscopic follow-up with the SED Machine (SEDM) mounted on the Palomar 60-inch telescope (P60) under standard BTS protocols.  The SEDM spectrum, obtained on August 21 \cite{2022TNSCR2465....1S}, shows an excellent match to a normal Type Ia supernova at a redshift of $z=0.35$ close to maximum light. The redshift and spectral classification were confirmed with a higher resolution spectrum obtained at the Nordic Optical Telescope on La Palma on the following night. We followed SN Zwicky up with P48 in the $g$ and $r$ bands. For our analysis we use the forced photometry as provided by the Image Processing and Analysis Center (IPAC) as detailed in \cite{2019PASP..131a8003M}. Observations in the SDSS $g$, $r$, $i$ and $z$ filters were taken with the IO:O optical imager on the Liverpool Telescope (LT) \cite{Steele2004}. The LT photometric data is processed with custom data-reduction and image subtraction software (Taggart et al 2022 in prep.). Image subtraction is performed using PS1 references image. Images were stacked using SWarp to combine multiple exposures where required. The photometry is measured using a point spread function (PSF) fitting methodology relative to PS1 standards and is based on techniques in \cite{Fremling2016}.

\subsection*{Lightcurve fit, magnification and Time Delay Inference}


{
We used the publicly available, python-based software \texttt{sntd} \cite{Pierel2019} for inferring the restframe $B$-peak magnitude, the lightcurve shape and colour \snia SALT2 parameters \cite{guy2010} and the time-delays between the \sn images.
Data points with $\geq 3 \sigma$ detections from the $g$ and $r$ filters from P48 and the $g,r,i,z$ filters from the Liverpool telescope were included in the fit. We adopted an iterative procedure in two steps. First, all the data was used. Next,  only data points in the range where the SALT2 model is defined, i.e. $-20$ to $+50$ days were kept for the second iteration. The final lightcurve fit  parameters were derived from data in this phase range. 

We estimated the time delays, i.e., the relative phase between the \sn images,  by fitting the unresolved ground-based flux data with the model which includes the flux contributions from the four sets \sn lightcurves, F$^j(t, \lambda)$, each one with their own time of $B$-band max, $t_0^{j}$. The fit is constrained by imposing a prior on the image ratios at the date of the Keck/NIRC2 observations, reported in Supplementary Table 3. 
We let the lightcurve fit parameters vary withing the ranges $x_1=[-3.0,3.0],c=[-0.3,0.3]$, but assume them to be the same for all four images. This is an excellent approximation in the absence of appreciable differential reddening in the lensing galaxy, confirmed by the result of the fit,
$c=-0.01 \pm 0.01$, consistent with no colour excess. }

Galactic extinction was included in the model, adopting $E(B-V)_{\rm MW} = 0.1558$ mag,  based on the extinction maps in \cite{Schlafly2011}. We used the wavelength dependence from the dust from \cite{cardelli1989} and the measured mean value of total to the Galactic selection extinction ratio, $R_V = 3.1$.  From the location of the fitted restframe $B$-band peak luminosity in the SALT2 model for the summed fluxes  we inferred the time of maximum for SN Zwicky, $t_0 = 59808.67 \pm 0.19$, corresponding to August 17, 2022. The total lensing magnification, $\mu = 24.3 \pm 2.7$ was calculated after standardising the \snia fitted $B$-band peak magnitude with the standard SALT2 lightcurve shape and colour correction parameters, ($\alpha$, $\beta$) = (0.14, 3.1). 
{Throughout the analysis of SN Zwicky, we adopted a flat $\Lambda$CDM cosmological model with $H_0=67.4$ km s$^{-1}$Mpc$^{-1}$ and $\om=0.315$ \cite{Planck2020:cosmo}.}


The uncertainties we report for the time delays account for parameter degeneracies in the the fit. Corner plots with posteriors for the relative time delays of B, C, and D with respect to A are illustrated in  Supplementary Fig. 1. The unresolved data used for this analysis will be available on WiseRep after paper acceptance. It should be emphasised that the lightcurve and time-delay fits were done completely independently from the lens modelling.

\subsection*{Spectroscopic follow-up}

The first classification spectrum of \Zw was obtained with the IFU on the SEDM on
2022 August~21. 
The data were reduced using a custom IFU pipeline 
developed for the instrument \cite{2019A&A...627A.115R,2022PASP..134b4505K}.   Flux calibration and correction of telluric bands were done 
using a standard star taken at a similar airmass.  The details of all spectroscopic observations are listed in Table~\ref{tb:speclog}.

We obtained three epochs of spectroscopy between 21 August and 11 September 2022 with the Alhambra Faint Object Spectrograph and Camera 
(ALFOSC) 
on the 2.56\,m Nordic Optical Telescope (NOT) at the Observatorio del Roque de los Muchachos on La Palma (Spain), {(see {http://www.not.iac.es/instruments/alfosc} for further information).}  Observations were taken using  grism 4, providing wavelength coverage over most of the optical spectral range (typically 3700--9600 \AA). Reduction and calibration were performed using PypeIt version 1.8.1 \cite{Prochaska2020b,Prochaska2020a}.

We obtained three epochs of spectroscopy between 22 August and  19 October 2022 with the Low Resolution Imaging Spectrometer (LRIS) on the Keck I 10 m telescope \cite{1995PASP..107..375O}.
All spectra were reduced and extracted with LPipe \cite{LPipe}.

\subsection*{Adaptive Optics observations }
\subsubsection*{Observations with the VLT} 
\paragraph{HAWKI observations and data reduction:}
We obtained 7 epochs of near-IR in $YJHK$ between 23 August and 30 September 2022 with the High Acuity Wide field K-band Imager (HAWK-I) \cite{Pirard2004a,Casali2006a,Kissler-Patig2008a} at the ESO Very Large Telescope at the Paranal Observatory (Chile). All observations, except of the first epoch, were performed with the ground layer adaptive optics offered by the GALACSI module \cite{Arsenault2008a,Paufique2010a,Stroebele2012A} to improve the image quality. The first three were observed in $YJH$ filters and the following 4 in $YJHK_s$ filters. For the first three epochs we exposed for $3\times60$~s in the $Y$ and $J$ bands and $6\times 60$~s in $H$ band. For the fourth epoch we also observe for $10\times60$~s in the $K_s$ band. To account for the brightness evolution, we exposed for $10\times100$~s and $6\times60$~s for epochs five and six. For the final epoch we also increased $H$-band exposure times to $10\times60$~s. 
For the HAWK-I observations we used offsets of ($ \ang{;;-115},  \ang{;;115}$) to place the target on the optimal detector chip.

The data used in our work has been reduced using the HAWK-I pipeline version 2.4.11 and the Reflex environment \cite{Freudling2013a}. The data reduction included subtracting bias and flat fielding. The world coordinate system was calibrated against stars from GAIA.

\paragraph{MUSE observations and data reduction:}

We obtained 4 epochs of integral-field spectroscopy between 24 August and 30 September 2022 with the Multi-Unit Spectroscopic Explorer (MUSE) \cite{Bacon2010a} at the ESO Very Large Telescope at the Paranal Observatory (Chile). Each pointing has an approximately $1'\times1'$ field of view with spatial sampling of $0.2''$/pixel and covers the wavelength range from 4700 to 9300~\AA\ with a spectral resolution of 1800 to 3600. All observations were performed with the ground layer adaptive optics offered by the GALACSI module \cite{Arsenault2008a,Stroebele2012A} to improve the image quality. The integration time of each epoch was 1800 s.

The data used in our work has been reduced using the MUSE
pipeline version 2.8.7 \cite{Weilbacher2014a} and the Reflex environment \cite{Freudling2013a}. The data reduction included subtracting bias, flat fielding, wavelength calibration and flux calibration against spectrophotometric standard stars. Afterwards, we improved the sky-subtraction with the Zurich Atmosphere Purge (ZAP) \cite{Soto2016a} module in the ESO MUSE pipeline. The world coordinate system was calibrated against stars from GAIA.

\subsubsection*{Laser Guide Star Adaptive Optics imaging from Keck}\label{sec:keckao}
The NIRC2 J-band observations consist of 9 images in a 5-point dither pattern based on a $\ang{;;2} \times \ang{;;2}$ grid size, to facilitate sky background subtraction. At the first dither location we obtained 2 images; one co-added 600s exposure and one 200s exposure. At the second location we obtained one 200s exposure. At each of the third, fourth and fifth dither locations, we obtained two 200s exposures. To correct for flat-fielding and bias we acquired a set of ten bias frames (flat lamp off) and ten dome flat (flat lamp on) frames.  Sky background and dark current was removed as part of the sky subtraction, which utilized a different sky map for each dither position, created by median combining the frames from all other dither positions, excluding the current dither position. The final combined image was created by aligning each dither position to each other using the centroid of the brightest SN image (Image A), and median combining. The reduction was carried out using custom python scripts. The NIRC2 $J$-band data (Fig.~2) provides the highest resolution (FWHM \ang{;;0.086}) image in our dataset of the system where the four SN images are visible.


\subsection*{Modelling of the lens galaxy}

The Keck NIRC2 $J$-band image was used to model the lens galaxy in terms of its Einstein radius $\te$, semi-minor to semi-major axis ratio $q$ (or ellipticity $\epsilon_e = 1 - q$) and orientation angle $\phi$. The mass profile used in our analysis is a singular isothermal ellipsoid (SIE) \cite{1993LIACo..31..571K,1994A&A...284..285K} :
\be
\label{eq:SIE}
\kappa(x, y) = \frac{\theta_\textrm{E}}{2 \sqrt{q x^2 + y^2/q}}, 
\ee
where $\kappa$ corresponds to the convergence (i.e. the dimensionless projected surface mass density) and the coordinates ($x, y$) are centred on the position of the lens centre and rotated counterclockwise by $\phi$. The projected mass $M$ inside an isodensity contour of the SIE is given by: \cite{1994A&A...284..285K}
\be
\label{eq:mass}
M=\frac{c^2}{4G}\frac{D_{\rm s}D_{\rm l}}{D_{\rm ls}} \te^2, 
\ee
where $D_{\rm l}$, $D_{\rm s}$, and $D_{\rm ls}$ are the angular diameter distances between the observer and the lens, the observer and the source, and the lens and the source, respectively. In order to calculate these distances, we assumed a flat $\Lambda$CDM cosmology with $H_0=67.4$ km s$^{-1}$Mpc$^{-1}$ and $\om=0.315$ \cite{Planck2020:cosmo}. 

In addition to the lens mass model, we included light models for the lens galaxy and SN host galaxy in the form of elliptical Sérsic profiles:
\be
I(R) = I_{\rm e}\exp{\left\{-b_n\left[\left(\frac{R}{R_{\rm e}}\right)^{1/n}-1\right]\right\}},
\ee
where $I_{\rm e}$ is the intensity at the half-light radius $R_{\rm e}$, $b_n=1.9992 n -0.3271$ \cite{birrer2018lenstronomy} and
\be
R\equiv\sqrt{x^2 + y^2/q_{\textrm{S}}^2},
\ee
with $q_{\textrm{S}}$ the axis ratio of the Sérsic profile. The SN images were modelled as point sources. 
{We used a Moffat PSF with power index 2.94 and FWHM \ang{;;0.091} to model the full image}. We simultaneously reconstructed the lens mass model, SN images, and the surface brightness distributions of the lens and host galaxy. {The lens mass model is constrained only by the positions of the lensed SN images and not by their fluxes, since the latter may be considerably affected by substructures, such as stars, in the lensing galaxy. Our model contains 13 non-linear free parameters: $\theta_{\rm E}$, $\phi$, $q$, $x$, $y$ for the lens mass model, $R_{\rm e}$, $n$, $\phi_{\textrm{S}}$, $q_{\textrm{S}}$, $x_{\textrm{S}}$, $y_{\textrm{S}}$ for the lens light model and $x_{\textrm{SN}}$, $y_{\textrm{SN}}$ for the SN position in the source plane. Our results are obtained using} \textsc{lenstronomy} {({https://lenstronomy.readthedocs.io/en/latest/})}, an open-source python package that uses forward modeling to reconstruct strong gravitational lenses \cite{birrer2018lenstronomy}. The result of the fit and comparison with the observations is shown in Supplemental Fig.~2.
As a cross-check, we also modeled the HST photometry data and redid the analysis with {\textsc{lensmodel}} \cite{Keeton:2001sr,Keeton:2001ss}, which produced consistent results.

The resulting best-fit values for the lens mass and light profiles are summarised in  Supplemental Table~2.
Additionally, we derived the gravitational mass within the isodensity contour given by the critical line of the lens (with radius $\te=0.167$ arcsec, corresponding to $0.628$ kpc) to be $M=(7.82\pm 0.06)\cdot 10^9\,M_\odot$.

Supplemental Table~3 displays the observed time delays and individual fractional flux contributions from each SN image as detailed in the subsection \textit{Lightcurve fit and Time Delay Inference} ($t_{\rm obs}$ and $f_{\rm obs}$) compared to the predictions from the lens model ($t_{\rm mod}$ and $f_{\rm mod}$). Here, time delays are given with respect to image A, e.g. $t_i\equiv t_i-t_{\rm A}$. 
The observed fractional flux ratios are measured from the Keck $J$-band image after subtracting the lens galaxy light, and the model predictions, $f_{\rm mod}$, are computed from the magnifications predicted by the lens model, $f_i\equiv \mu_i/\sum_j\mu_j$. In addition to the uncertainties obtained from the $J$-band image analysis, we make a conservative error estimate by also including the scatter in $f_{\rm obs}$ and $f_{\rm mod}$ obtained when modelling the system using data from the HST optical filters $F475W$, $F625W$ and $F814W$, as well as the Keck near-IR $J$-band data. Using this approach, we also take into account possible error contributions from uncertainties in dust extinction, lens galaxy subtraction, and lens mass modelling.

The observed individual image magnifications can be obtained from $f_{\textrm{obs}}$ by multiplying the individual fractional fluxes with the total observed SN magnification of $\mu^{\rm tot}_{\rm obs}=24.3\pm 2.7$. 
The total lens model image magnification, $\mu^{\rm tot}_{\rm mod}$, is sensitive to the lens mass slope (which is unconstrained by the imaging data), such that flatter halos predicts a larger magnification. {For an isothermal lens, $\mu^{\rm tot}_{\rm mod}=14.9\pm 0.9$, indicating a flatter halo profile} (see also \cite{2017Sci...356..291G}). However, the predicted flux ratios remain approximately constant, which means that to first approximation, we can multiply the derived $f_{\rm mod}$ with an arbitrary $\mu^{\rm tot}_{\rm mod}$. In contrast to the predicted individual fractional flux contributions, the observed flux is dominated by images A and C. In order to match the observed values, substructure lensing is needed to additionally magnify images A and C with factors of $>4$ and $>2$, respectively, compared to any additional substructure (de)magnifications of images B and D.  

{In order to check the impact of added macro lens model complexity, we have studied cases where the lens mass distribution is modelled with two matter components; one where the surface mass density follows the lens light distribution (with an arbitrary mass to light ratio, possibly interpreted as a baryonic mass component), and a second one introducing a dark matter halo with additional flexibility on density profiles. In spite of the added extra complexity, the quality of the fit to the \sn image positions does not improve, and only induces shifts in the predicted flux ratios below 5 \%, i.e., very small in comparison with the observed flux ratio anomalies.} 

{We do not detect any time dependence in the flux ratios between the Keck and HST images, observed just one month after lightcurve peak, six days apart, nor any other anomalous variation in the unresolved $\sim80$ days long lightcurve shown in Fig.~\ref{fig:lc}. Comparing the expected size of SN photosphere with the Einstein radii of compact objects in the lensing galaxy, we infer that if the discrepancy from the smooth lens model is caused by microlensing, the deflectors must exceed 0.02 $M_\odot$.}  

With the aim of distinguishing between lensing effects by stars or larger substructures, we investigated the upper limit of image splitting for the brightest image, A. We approximated the maximum Einstein radius of a large structure at the position of image A by putting an upper limit on the difference in the full width at half maximum between $\textrm{PSF}_{\textrm{A}}$ and $\textrm{PSF}_{\textrm{BCD}}$. Using equation~\ref{eq:mass}, we inferred a $95\%$ confidence upper limit on the substructure's mass within its Einstein radius. {Combined with the lower mass limit derived from the lack of flux ratio variability, this constrains the mass of the substructure deflector in the line-of-sight to image A to $0.02 < M/M_\odot < 3\cdot 10^7$.}

We conclude that, similarly to the case of iPTF16geu, a smooth lens density fails to explain the individual image magnitudes and additional sub-structure lensing is needed.
Since the observed properties of lens systems to first order only depend on the integrated mass within the images and/or the surface mass density of the lens at the image positions or in the annulus between the images \cite{Kochanek:2004ua}, small image separation systems provide a unique probe of the central regions of gravitational lensing galaxies. 
The probability for lens substructures, such as stars, to accommodate the needed (de)magnifications for a range of lens density slopes will be investigated in a future lens modeling paper.

\subsection*{Lens galaxy photometry}\label{sec:lens_galaxy}

We retrieved science-ready coadded images from the Panoramic Survey Telescope and Rapid Response System (PS) DR1 \cite{Chambers2016a}. Using a circular aperture with a radius of \ang{;;1.1}, we obtain the following apparent magnitudes of the lens galaxy: $g=22.09\pm0.09$, $r=20.71\pm0.02$, $i=20.14\pm0.02$, $z=19.84\pm0.02$, and $y=19.63\pm0.05$ (all errors are of statistical nature). We model the spectral energy distribution with the software package prospector version 1.1 \cite{Johnson2021a}. Prospector uses the Flexible Stellar Population Synthesis (FSPS) code \cite{Conroy2009a} to generate the underlying physical model and python-fsps \cite{ForemanMackey2014a} to interface with FSPS in python. The FSPS code also accounts for the contribution from the diffuse gas (e.g., H II regions) based on the Cloudy models from \cite{Byler2017a}. Furthermore, we assumed a Chabrier initial mass function \cite{Chabrier2003a} and approximated the star formation history (SFH) by a linearly increasing SFH at early times followed by an exponential decline at late times (functional form $t \times \exp\left(-t/\tau\right)$). The model was attenuated with the \cite{Calzetti2000a} model. 

The best fit galaxy model points to a moderately massive galaxy with stellar mass $\log\,M_\star/M_\odot=10.1\pm0.3$. The other parameters, such as star-formation rate, attenuation and age are poorly constrained and we report their values only for the sake of completeness: ${\rm SFR}=15^{+21}_{-15}~M_\odot\,\rm yr^{-1}$, $E(B-V)_{\rm  star}=0.6^{+0.1}_{-0.4}$ mag, $\rm Age=1.6^{+5.4}_{-1.2}~\rm Gyr$. We acknowledge that the aperture might not encircle the entire galaxy and, therefore, we might underestimate the stellar mass of the lens. Changing the radius of the measurement aperture to \ang{;;4} increases the mass by $\approx0.3$~dex. This large aperture also includes the contribution of the SN host galaxy and therefore overestimate the stellar mass of the lens galaxy. Nonetheless, this upper bound does not alter our conclusion about the compact nature of the lensing galaxy, relative to other lensing systems.

\subsection*{Host galaxy photometry}

Numerous studies have shown that the peak absolute magnitudes of SNe Ia depend on their host galaxy masses, see e.g., \cite{2010MNRAS.406..782S}. Although the host and the lens galaxy are well separated in the HST images, both galaxies have diffuse emission extending well beyond the separation of two galaxies. To first order, we can remove the contribution of the lens by subtracting the lens images predicted by \textsc{lenstronomy} from the HST images. Using a circular aperture with a \ang{;;1} radius and appropriate zeropoints from HST, we measure for the host: $22.31 \pm 0.11$, $21.81 \pm 0.06$ and $21.13 \pm 0.06$ mag in $F475W$, $F625W$ and $F814W$, respectively. Fitting the SED with prospector with the same assumptions as in the previous Section gives a galaxy stellar mass of $\log M_\star/M_\odot = 9.6^{+0.4}_{-0.3}$. A closer inspection of the subtracted HST images reveals that the \ang{;;1} aperture does not encircle the total emission of the host. Increasing the aperture radius to \ang{;;1.5} encircles almost the entire host galaxy ($F475W=21.80 \pm 0.09$, $F652W=21.38 \pm 0.04$, $F814W=20.74 \pm 0.03$), but it also includes non-negligible contribution from residuals of the lens galaxy. The galaxy mass increases marginally to $\log M/M_\odot = 9.7^{+0.5}_{-0.4}$ but is consistent with the previous measurement. This suggest that a mass-step correction is not required. 
To get a more robust estimate of the galaxy mass, NIR observations after the SN has faded are required.

\section*{Data Availability}
The reduced spectra and lightcurves used in the paper are available through the WISEREP repository \cite{2012PASP..124..668Y} (https://www.wiserep.org/object/21343).   The raw VLT data will be also available from the ESO Science Archive Facility. 
\section*{Code Availability}
The data was analysed using the public codes SNTD ({https://sntd.readthedocs.io/en/latest/}) and LENSTRONOMY ({https://lenstronomy.readthedocs.io/en/latest/}). The Keck AO J-band images can be requested from the first author.


\section*{Acknowledgements} 
Based on observations obtained with the Samuel Oschin Telescope 48-inch and the 60-inch Telescope at the Palomar Observatory as part of the Zwicky Transient Facility project. ZTF is supported by the National Science Foundation under Grant No. AST-2034437 and a collaboration including Caltech, IPAC, the Weizmann Institute of Science, the Oskar Klein Center at Stockholm University, the University of Maryland, Deutsches Elektronen-Synchrotron and Humboldt University, the TANGO Consortium of Taiwan, the University of Wisconsin at Milwaukee, Trinity College Dublin, Lawrence Livermore National Laboratories, IN2P3, University of Warwick, Ruhr University Bochum and Northwestern University. Operations are conducted by COO, IPAC, and UW. SED Machine is based upon work supported by the National Science Foundation under Grant No. 1106171. 
This work has been supported by the research project grant "Understanding the Dynamic Universe" funded by the Knut and Alice Wallenberg Foundation under Dnr KAW 2018.0067, and the G.R.E.A.T research environment, funded by {\em Vetenskapsr\aa det}, the Swedish Research Council, project number 2016-06012. MR acknowledges support from the European Research Council (ERC) under the European Union's Horizon 2020 research and innovation programme (grant agreement n$^o$759194 - USNAC). 
AG acknowledges support from the Swedish Research Council under contract 2020-03444. EM acknowledges support from the Swedish Research Council under contract 2020-03384. T.E.C. is funded by a Royal Society University Research Fellowship and the European Research Council (ERC) under the European Union's Horizon2020 research and innovation program(LensEra: grant agreement No. 945536).
Based on observations collected at the European Organisation for Astronomical Research in the Southern Hemisphere under ESO programmes 109.234A.001 and 109.24FN. Based on observations made with the Nordic Optical Telescope, owned in collaboration by the University of Turku and Aarhus University, and operated jointly by Aarhus University, the University of Turku and the University of Oslo, representing Denmark, Finland and Norway, the University of Iceland and Stockholm University at the Observatorio del Roque de los Muchachos, La Palma, Spain, of the Instituto de Astrofisica de Canarias. Some of the data presented here were obtained in part with ALFOSC, which is provided by the Instituto de Astrofisica de Andalucia (IAA) under a joint agreement with the University of Copenhagen and NOT.
Some of the data presented herein were obtained at the W. M. Keck Observatory, which is operated as a scientific partnership among the California Institute of Technology, the University of California and the National Aeronautics and Space Administration. The Observatory was made possible by the generous financial support of the W. M. Keck Foundation. The Liverpool Telescope is operated on the island of La Palma by Liverpool John Moores University in the Spanish Observatorio del Roque de los Muchachos of the Instituto de Astrofisica de Canarias with financial support from the UK Science and Technology Facilities Council. This paper is based in part on observations with the NASA/ESA Hubble Space Telescope obtained from the Mikulski Archive for Space Telescopes at STScI; support was provided to JP through program HST-GO-16264.

\section*{Author Contribution Statement}
AG: Project lead, telescope proposals, main manuscript editor; JJ, SS,NA,ASC, SD, EM: Data analysis, proposal contribution, figures, manuscript text; CF, LY, WM: Keck AO imaging, 
DP, KH: Liverpool telescope observations; JS: Nordic Optical Telescope observations; RJ, IA: data analysis; TC, CL: manuscript contribution; EB, JB,MG, MK, SK, AM, JDN, JN, KN, RR, MR, BR, GS, AT, AW: ZTF observations; YS, RS: Keck spectroscopy; JV, JCW: follow-up observations; JP; HST observations. J.R., Very Large Telescope observations.

\paragraph{Correspondence:} Ariel Goobar, ariel@fysik.su.se 

\section*{Competing Interest Statement}
The authors declare no competing interests.

\newpage




\setcounter{page}{1}
\section*{Supplementary Information} 

\noindent Tables S1, S2, S3\\
\noindent Figures S1, S2 \\
\begin{table}[!htb]
\begin{tabular}{lcccccc}
\hline\hline
UT Date &  MJD   & Phase & Telescope  & $R$ & 
$\lambda$ range & Exp. time   \\ 
& (days) & (days) & Instrument &  ($\lambda / \Delta \lambda$) & (\AA) & (s) \\ \hline
2022-08-21.21 & 59812.21 & +2.7 & P60/SEDM & 100 & 4000-9200 & 2250 \\ 
2022-08-22.92 & 59813.92 & +3.9 & NOT/ALFOSC & 360 & 3700-9200 & 2400 \\ 
2022-08-23.28 & 59814.28 & +4.2 & Keck/LRIS	& 820 & 3200 - 10200 & 600 \\ 
2022-09-30.03 & 59852.03 & +32.1 & VLT/MUSE & 1800 - 3800 & 4700-9300 & 1800  \\	
\hline
\end{tabular}
\caption{%
{\bf Spectroscopic observations of SN Zwicky used for classification and redshift measurements.}}
\end{table}
\begin{table*}[h]
  \begin{tabular}{cccccc}
    \hline \hline
    & $\te$ [$''$] & $\phi$ [$^\circ$] & $q$ & {$R_{\rm e}$} & {$n$} \\
    \hline 
    Lens mass & $0.1670\pm 0.0006$ & $68.87\pm 0.15$ & $0.653\pm 0.010$ & - & - \\
    Lens light & - & $68.11\pm 0.74$ & $0.661\pm 0.008$ &  {$1.74\pm 0.15$}  & {$2.73\pm 0.10$} \\
    \hline    
  \end{tabular}
 \caption{\bf{Summary of lens mass and light constraints from the Keck $J$-band data (Einstein radius $\te$, orientation angle $\phi$, axis ratio $q$, {half-light radius $R_{\rm e}$, and Sérsic index $n$}). Note the agreement between the lens mass and light distributions. }}
  \label{tab:lenstronomy}
\end{table*}

\begin{table}
  
  \begin{tabular}{cccccc}
    \hline \hline
    Image & $\Delta t_{\rm obs}$ [days]& $\Delta t_{\rm mod}$ [days]& $f_{\rm obs}$  & $f_{\rm mod}$ & {$f_{\rm fit}$} \\
    \hline
    A & -               & -                & $0.367\pm 0.013$ & $0.150\pm 0.004$ & {$0.376\pm 0.084$}\\
    B & $-0.4 \pm 2.9$  & $-0.51 \pm 0.02$ & $0.154\pm 0.005$ & $0.285\pm 0.003$ & {$0.149\pm 0.060$}\\
    C & $-0.1 \pm 2.3$  & $-0.27 \pm 0.01$ & $0.304\pm 0.011$ & $0.257\pm 0.004$ & {$0.300\pm 0.072$}\\
    D & $-0.1 \pm 2.7$  & $-0.44 \pm 0.02$ & $0.175\pm 0.005$ & $0.309\pm 0.004$ & {$0.174\pm 0.065$}\\
    \hline
  \end{tabular}
  \caption{{\bf Observed arrival time differences (derived from the light curve) and individual fractional flux contributions from each SN image, compared to the ones predicted by the lens model {and the lightcurve modelling}. The discrepancy between the individual flux contributions shows that a smooth lens model cannot explain the observed SN image fluxes.}}
  \label{tab:obsvsmodel}
\end{table}

\begin{figure}
    \centering
    \includegraphics[width=1.0\textwidth]{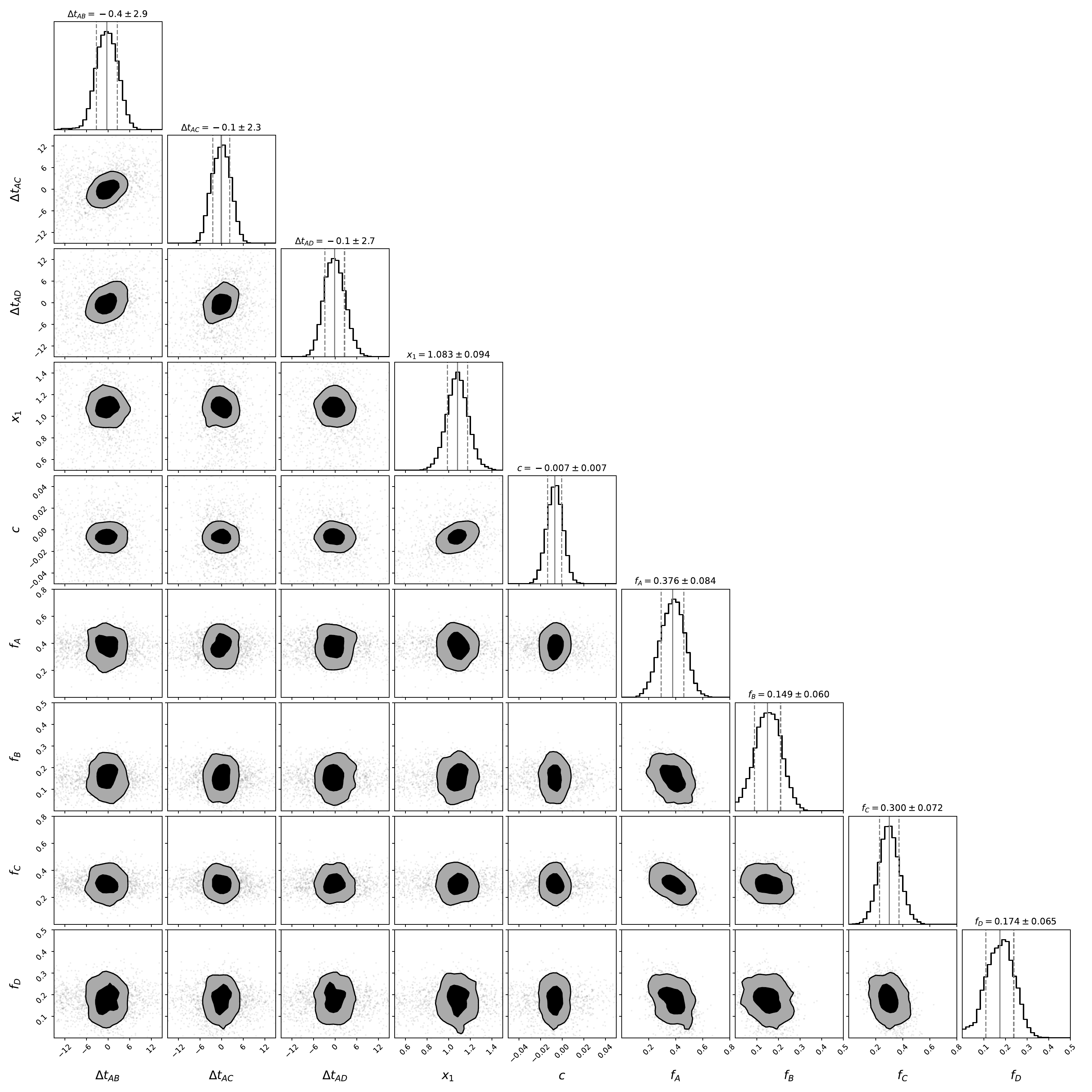}
    \caption{{\bf Corner plot showing the posterior distribution for the relative time-delay of the individual images B, C, and D to A,{the fitted supernova parameters $x_1$ and $c$, as well as the reconstructed flux ratios at peak from the fit} with the relevant correlations between each parameter. The 1D plots show the mean value with the 1 sigma uncertainty. The 2D plots show the 1 and 2 sigma contours.}}
    \label{fig:cornerplot}
\end{figure}

\begin{figure*}
    \centering
	\includegraphics[width=1\textwidth]{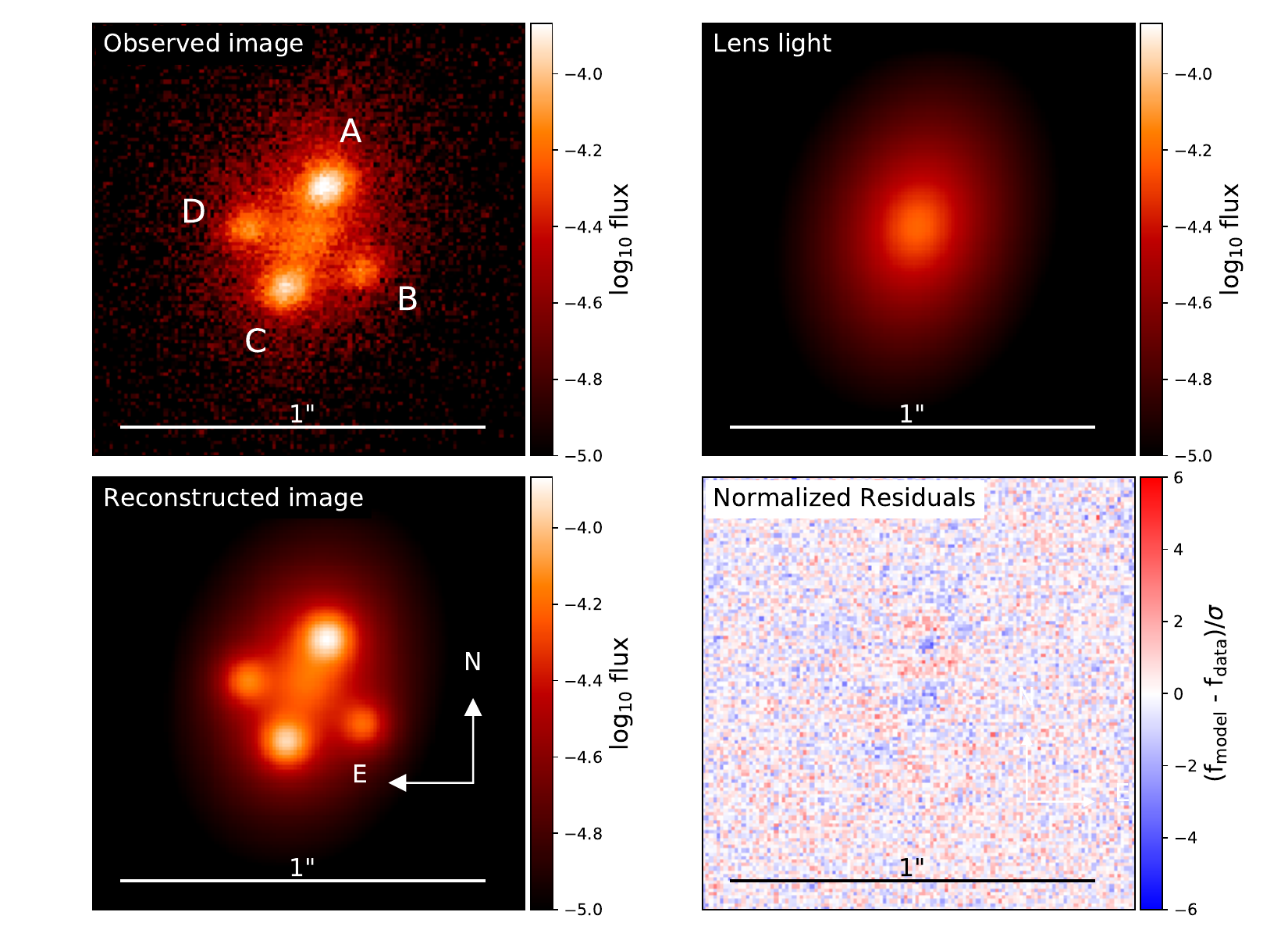}
	\caption{ {\bf A comparison between the observed (data) image and the reconstructed (model) image for the Keck $J$-band photometry. {\it Upper panel from left to right:} The observed image and the reconstructed lens galaxy light intensity. {\it Lower panel from left to right:} The total reconstructed image (convolved with the PSF of the instrument, but without background or Poisson noise) and the normalized residuals. 
	\label{fig:J_SN_g20}}}
\end{figure*}
\clearpage
\bibliography{scibib}

\end{document}